\documentclass[aps,superscriptaddress,onecolumn,notitlepage]{revtex4-1}
\usepackage{graphicx}
\usepackage{bm}
\usepackage{amsmath}

\newcommand{\bra}{\langle}
\newcommand{\ket}{\rangle}

\newcommand{\bs}[1]{\ensuremath{\boldsymbol{#1}}}

\newcommand{\be}{\begin{equation}}
\newcommand{\ee}{\end{equation}}
\newcommand{\bea}{\begin{align}}
\newcommand{\eea}{\end{align}}
\newcommand{\beqa}{\begin{eqnarray}}
\newcommand{\eeqa}{\end{eqnarray}}
\newcommand{\ev}[1] {\left|\left \langle #1  \right \rangle \right|^2}

\begin{document}

\title{Log-Periodic Oscillations in the Photo Response of Efimov Trimers}

\date{\today}

\author{Betzalel Bazak}
\email{betzalel.bazak@mail.huji.ac.il}
\affiliation{Racah Institute of Physics, The Hebrew University, 
             91904, Jerusalem, Israel}

\author{Nir Barnea}
\email{nir@phys.huji.ac.il}
\affiliation{Racah Institute of Physics, The Hebrew University, 
             91904, Jerusalem, Israel}

\begin{abstract}
The photoassociation of Efimov trimer, composed of three identical bosons,
is studied utilizing the multipole expansion. 
For identical particles the leading contribution comes from the 
$r^2$ $s$-mode operator and from the quadrupole $d$-mode operator. 
Log-periodic oscillations are found in the photoassociation response function,
both near the energy threshold for the leading $s$-wave reaction, 
and in the high frequency tail for all partial waves.
\keywords{Photo reactions \and Efimov physics \and Ultracold gases}
\end{abstract}
\maketitle

\section{Introduction}
\label{intro}
The implementation of photo-association techniques in ultracold atomic traps 
\cite{ThoHodWie05} opened a new route for quantitative  
determination of universal properties in few-body systems \cite{BraHam06}.
In these experiments, radio-frequency (rf) induced trimer formation leads 
to enhanced atom loss rates from the traps.
Scanning the rf field frequency, the change in the measured atom loss rate 
indicates various molecular thresholds and structures.

Recently, trimer formation through rf association was discovered in both 
fermionic $^6$Li \cite{LomOttSer10, NakHorMuk11}
and bosonic $^7$Li \cite{MacShoGro12} systems.
The three body case attracts special attention as the simplest non-trivial 
system. Moreover, in the 70's Efimov predicted that in the limit of a 
resonant 2-body interaction, the system reveals universal properties 
\cite{Efi70}.
A peculiar prediction is the existence of a series of giant three body 
molecules, known as Efimov trimers, that was verified experimentally 
few years ago \cite{ZacDeiDer09,PolDriHul09,GroShoKok09,GroShoMac11}.

In a previous work \cite{BazLivBar12}, 
we have presented the multipole analysis of an rf association process 
binding a molecule of $N$ identical bosons.
We have shown that the spin-flip and frozen-spin processes differ by 
their operator structure and by the de-excitation modes that contribute 
to the photoassociation rate. 
Previous analysis of these rf experiments 
\cite{ChiJul05, BerJesMol06, HanKohBur07, KleHenTop08,TscRit11},
which relied on the Franck-Condon factor, 
is appropriate for describing spin-flip reactions. 
For frozen-spin reactions we have applied our results to study 
the dimer formation \cite{BazLivBar12}, and to study numerically 
the quadrupole response of a bound bosonic trimer  
\cite{LivBazBar12,BazLivBar13}. 

Here, we study trimer photoassociation using the hyperspherical adiabatic approximation. 
A zero-range potential is used to derive analytic results 
for the transition rates at the unitary point. 
Similarly to the dimer case \cite{BazLivBar12}, the $s$-mode and the $d$-mode 
are found to be the leading order contributions.  A new fingerprint of
Efimov physics is studied, which is a log-periodic oscillation in the response.

%=============================
\section{Multipole expansion}
%=============================

The molecular photoassosiaction rate is given by Fermi's golden rule,
\be \label{golden_rule}
r_{i\rightarrow f}=\frac{2\pi}{\hbar} \bar{\sum_i} \sum_f
|\bra f, \bs k \zeta | \hat{H}_I | i\ket|^2 
\delta(E_i-E_f-\hbar\omega_k) \;,
\ee 
where three particles in an initial continuum state with energy $E_i$ 
form a bound state with energy $E_f$
by emitting a photon with momentum $\bs k$, polarization $\zeta$ and energy 
$\hbar \omega_k=\hbar c k$.
$\bar\sum_i$ is an average on the appropriate initial continuum states 
and $\sum_f$ is a sum on the final bound states.
The coupling between the neutral atoms and the radiation field takes the form 
$\hat{H}_I=-e \int d\bs x \; \bs \mu(\bs x) \cdot \nabla \times \bs A(\bs x)$,
where $\bs A$ is the electromagnetic (EM) photon field, and $\mu(\bs x)$ is
the magnetization current. Here we consider 
only the one-body current
$\mu(\bs x)=\mu_0 \sum_j\bs S_j \delta(\bs x-\bs r_j)$,
where $\mu_0$ is the magnetic moment of a single particle, and 
$\bs S_j$ and $\bs r_j$ are the spin and position of particle $j$.

We assume that the initial and final atomic wave functions can be written as 
a product of symmetric spin $|\chi\ket$ and configuration space $|\psi\ket$ terms,
and that the photon does not induce change in the spin structure of the system. 
In this case the transition matrix element can be written as
\be \label{frozen-spin}
\bra f,\bs k \zeta | \hat H_I | i\ket = -i \mu_0 \sqrt{\frac{\hbar c^2}{2
    {\cal V}\omega_k}} k \bra S_{0}  \ket
\bra \psi^f_{L' M'} | \sum_{j=1}^3 e^{i\bs k \cdot \bs r_j} | \psi^i_{L M} \ket \;,
\ee
where 
$\bra S_{0} \ket = \frac{1}{3} \sum_j \bra \chi^i_{M_F} |S_{j,0}| \chi^i_{M_F}\ket $ 
is the average single particle magnetic moment, which plays the role 
of an effective charge. We normalize the EM field in a box of volume ${\cal V}$.

The photon wavelength of rf radiation is much larger than the typical 
dimension of the system $R$, therefore $kR\ll1$ and the lowest order 
in $kR$ dominates the interaction. 
Therefore the exponent can be expanded to get 
\be \label{expand}
\sum_{j=1}^3 e^{i\bs k \cdot \bs r_j} \approx  3
+i\sum_{j=1}^3 \bs k \cdot \bs r_j
-\frac{1}{6} \sum_{j=1}^3 k^2 r_j^2
-\frac{4\pi}{15} \sum_{j=1}^3 k^2 r_j^2 \sum_m Y_{2 -m}(\hat{k}) Y_{2 m}(\hat{r}_j), 
\ee
where $Y_{lm}$ are the spherical harmonics.
Each order in this expansion has clear physical meaning.
The zeroth order operator stands for elastic interaction.
The first order operator is the dipole, which for identical particles is
proportional to the center of mass and hence does not affect the relative 
motion of the atoms. 
At second order two operators appear: the $r^2$ operator, corresponding 
to $s$-mode reaction, and the quadrupole terms, corresponding 
to $d$-mode reaction.
For identical particles this is the leading term in low energy frozen spin reactions, 
and the transition probability scales as $k^5$.   
Summing over the initial and final magnetic numbers $M$, $M'$, 
the transition matrix element reads 
\be \label{t-frozen}
\sum_{M,M'}
| \bra f,\bs k \zeta | \hat H_I | i \ket |^2  = 
\frac{4\pi \hbar c k^5 \mu_0^2}{2\Omega} \ev{S_0}
\left( \frac{1}{6^2} 
| \bra \psi^f_{L'} \Vert \sum_{j=1}^3 r_j^2 Y_0\Vert \psi^i_{L} \ket |^2
+\frac{1}{15^2} 
| \bra \psi^f_{L'} \Vert \sum_{j=1}^3 r_j^2 Y_2(\hat r_j)  \Vert \psi^i_{L} \ket |^2
\right).
\ee

%===============================
\section{The Three Body Problem}
%===============================

The dynamics of a quantum 3 particle system is governed 
by the Schroedinger equation 
\be \label{Hamiltonian}
\left(T+W\right)\psi=E\psi
\ee
where $T$ is the center of mass kinetic energy operator and $W$ is the 
potential. In this study we shall limit our attention to
short range 2-body forces, thus $W=\sum_{i<j}V(|\bs{r}_i-\bs{r}_j|)$.
To eliminate the center of mass motion, we define the Jacobi coordinates,
$\bs x=\sqrt \frac {1}{2} (\bs r_2-\bs r_1)$, and $\bs y = 
\sqrt \frac {2}{3}\left(-\bs r_3 + \frac{\bs r_1+\bs r_2}{2}\right)$,
which we transform into the hyperspherical coordinates
$\rho^2=x^2+y^2$, and $\Omega=(\alpha,\hat x, \hat y)$, where 
$\tan \alpha=x/y$.

In the limit of infinite scattering length, $|a|\rightarrow\infty$,
the spatial wave function can be written as \cite{Mac68} 
$ \psi(\rho,\Omega)=\rho^{-5/2}\mathcal R(\rho)\Phi(\Omega) $.
The hyperspherical functions $\Phi(\Omega)$ and the corresponding 
eigenvalue $\nu^2$, are the solutions of the hyperangular equation,
\be \label{hae}
 \left ( \hat K ^2 + \frac{2m\rho^2}{\hbar^2}\sum_i 
V(\sqrt 2 \rho \sin \alpha_i) +4 \right)\Phi=
 \nu^2 \Phi, \ee
where
$
\hat K^2=-\frac{1}{\sin 2\alpha}\frac{\partial^2}{\partial \alpha^2}\sin2\alpha
+\frac{\hat l^2_x}{\sin^2\alpha}+\frac{\hat l^2_y}{\cos^2\alpha}-4.
$
For low energy physics, when the extension of the wave function is
much larger than the range of the potential, one can utilize the zero range
approximation. In this approximation the lateral extension of the potential is
neglected all together, and the action of the potential is represented through
the appropriate boundary conditions.
For a two-particle system the low energy interaction is dominated by the
$s$-wave scattering length $a$ and the wave function fulfills the boundary
condition $[u'/u]_{r=0}=-1/a$. The corresponding 3-body condition is
\be \label{bc}
\left[\frac{1}{2\alpha \Phi}\frac{\partial}{\partial \alpha}2 \alpha \Phi
\right]_{\alpha=0}=-\sqrt 2 \frac{\rho}{a}\;.
\ee

Plugging the solution of Eq. (\ref{hae}) into Eq. (\ref{bc}), one gets 
transcendental equations for $\nu$.
For $L=0$ the resulting equation is \cite{FedJen01}
\be\label{l0}
  \nu\cos(\nu\pi/2)-\frac{8}{\sqrt 3}\sin (\nu\pi/6)=\frac{\sqrt 2 \rho}{a}
  \sin(\nu\pi/2).
\ee
For $|a|=\infty$ the solution with lowest $\nu^2$ is $\nu_0\approx
1.00624i$, corresponding to the Efimov trimer. 

For $L=2$ the corresponding equation is,
\be \label{l2}
  \nu(4-\nu^2)\cos(\nu\pi/2)+
  24 \nu \cos(\nu \pi/6) +
  \frac{8}{\sqrt3} (\nu^2-10) \sin (\nu \pi/6)=
  -\frac{\rho}{a}(\nu^2-1)\sin(\nu\pi/2)
\ee
For $|a|=\infty$ the lowest non-trivial solution is
$\nu_2\approx2.82334$. 
 
%==========================
% The Hyperradial Equation 
%==========================

In the limit of $|a|\rightarrow\infty$, $\nu_L(\rho)=\nu_L$ and 
the hyperradial equation for $\mathcal R(\rho)$ is similar to the Bessel equation, 
\be\label{Bessel} 
-\mathcal R''(\rho)+\frac{\nu_L^2-1/4}{\rho^2} \mathcal R(\rho)
=\epsilon \mathcal R(\rho).
\ee
where $\epsilon=2mE/\hbar^2$.

We seek the solution for the bound ($\epsilon<0$) and continuum ($\epsilon>0$) cases:

I. A \emph{bound state} exists only for $L=0$, and $\nu_0\approx1.00624i$.
In this case the relevant solution is
$\sqrt \rho K_{\nu_0}(\kappa \rho)$, where $\kappa=\sqrt {-\epsilon}$. 
At the origin, this solution behaves like 
$\sin (\nu \ln (\kappa \rho/2)+0.301)$, 
therefore regularization is needed to avoid collapse, 
e.g. setting $\mathcal R(\rho \le \rho_0)=0$ for some finite $\rho_0$.
The result is the discrete Efimov spectrum,
$\epsilon_n/\epsilon_0=e^{-2\pi n/|\nu_0|}\approx 515^{-n}.$
The normalized wave functions are 
\be
\mathcal R^{(n)}_B(\rho)=\sqrt{2 \sin \nu_0 \pi/\nu_0 \pi} \kappa_n \sqrt \rho K_{\nu_0}(\kappa_n \rho)
\ee.

II. For \emph{continuum state}, the solution is 
\be
\mathcal R_L(\rho)=\sqrt{\frac{q \rho N_s}{2 R}} \left[ \sin \delta \mathrm{Re}[J_{\nu_L}(q \rho)] +
\cos \delta \mathrm{Re}[Y_{\nu_L}(q \rho)] \right]
\ee
where $q=\sqrt {\epsilon}$, $N_s=1/2(\pi)$ for imaginary (real) $\nu$, 
and we assume normalization in a sphere of radius $R$.
The phase shift $\delta$ is to be found from the boundary condition, 
$\mathcal R_L(\rho_0)=0$.

%=======================================
\section{The Transition Matrix Elements}
%=======================================

Now that we have obtained the initial and final wave functions
we are in position to evaluate the transition matrix elements, 
Eq. (\ref{t-frozen}), 
$I_\lambda=\bra \psi^f_{L'} \Vert \sum_{j=1}^N r_j^2 Y_\lambda \Vert \psi^i_L \ket$,
for $\lambda=s (d)$ corresponding to the $r^2$ (quadrupole) operator, respectively.

\begin{figure}
  \centering
  \includegraphics[trim=0cm 0cm 0cm 0cm, clip=true, height=5.4cm]{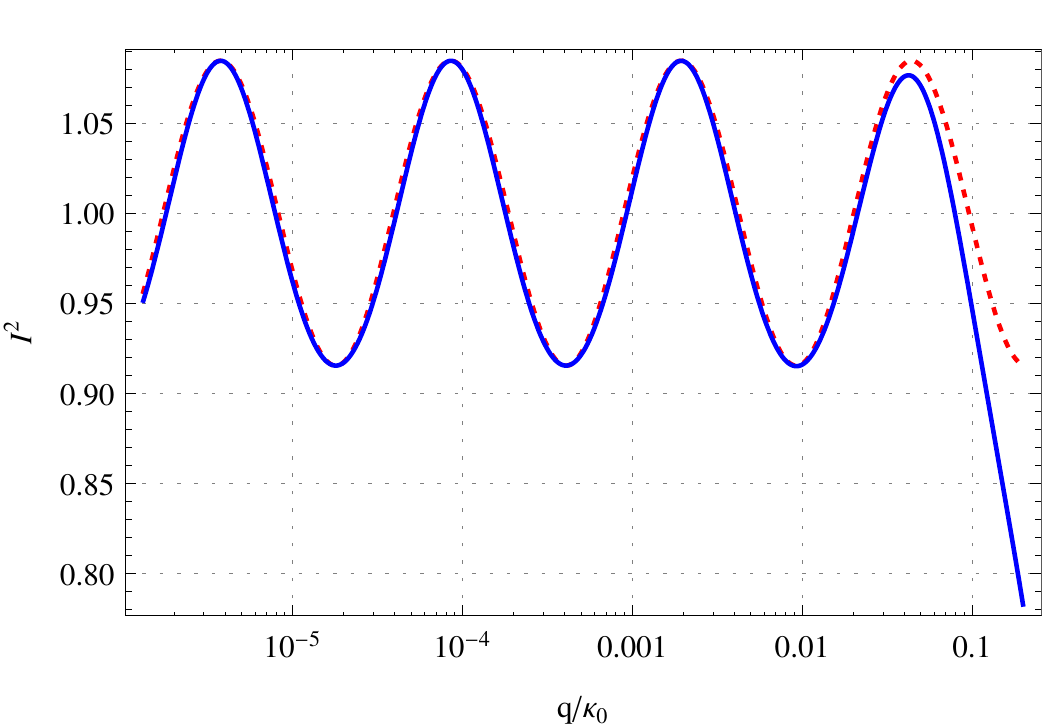}
  \includegraphics[trim=0cm 0cm 0cm 0cm, clip=true, height=5.4cm]{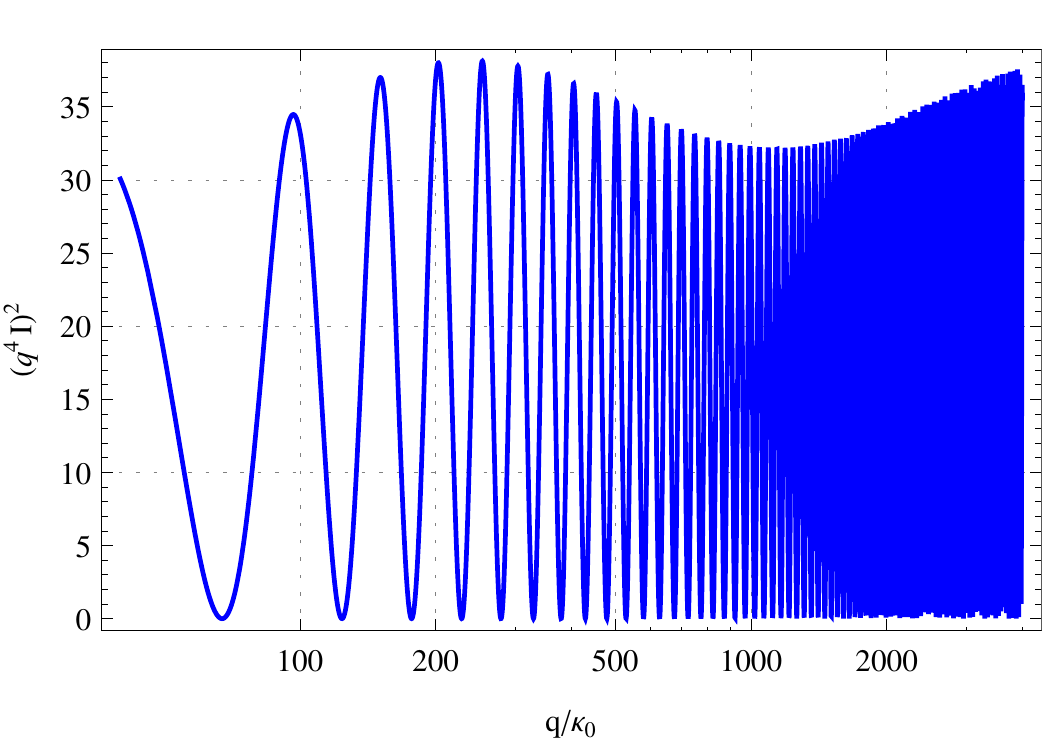}
  \caption{The log periodic oscillation in the $r^2$ transition matrix element.
	Left: near the threshold. Right: at the high frequency tail.}
  \label{fig:Osc} 
\end{figure}

\emph{The $r^2$ matrix element} ---
%----------------------------------
The $r^2$ operator connects the $L=0$ bound state to an $L=0$ scattering 
state. We note that $\sum_j r_j^2 = \rho^2 + 3R_{CM}^2$
where $R_{CM}$ is the center of mass radius. As we neglect the center of mass 
excitations, the matrix element is reduced into the hyperradial integral 
$ I_s(\kappa,q)=\int_0^\infty d\rho \mathcal R_B^*(\rho) \rho^2 \mathcal R_s(\rho)$.
Evaluating this integral, 
the resulting response function reveals log periodic oscillations \cite{BazBar13b}.
At threshold, the matrix element gets a particularly simple form which can be well approximated by
\be
I_s(\kappa,q) \approx C \kappa^{-3}\sqrt{q}\left(1+\frac{B_3}{2}\cos(2s_0\ln\frac{q}{\kappa})\right)
\ee
where $C$ is a constant that contains the normalization factors, and $B_3 \approx
8.475\%$ is the normalized amplitude of the oscillations.  
The oscillations modulate the matrix element all the way to the high energy tail. 
In Fig. \ref{fig:Osc} we present the 
log periodic oscillations of the $r^2$ matrix element.

\emph{The Quadrupole matrix element} ---
%-----------------------------------------
The quadrupole operator $\sum_j r_j^2 Y_2^M(\hat r_j)$ connects the $L=0$ 
bound state with $L=2$ scattering states. 
In this case the reduced matrix element takes the form
\be
I_d(\kappa,q) =
\frac{3}{2}\sqrt{5}
\int_0^\infty d\rho \mathcal R_B^*(\rho) \rho^2 \mathcal R_d(\rho)
\int d\Omega \Phi_f^*(\Omega) \sum_j \cos^2 \alpha_j Y_{20}(\hat y_j)\Phi_i(\Omega).
\ee
In this case we find no log-periodic oscillations near threshold.
Such oscillations, however,  modulate 
the high energy tail of the response function, 
attenuated by $q^{-4}$ and masked by the linear phase shift variation.
These high energy oscillations appear not only in these cases but in
all partial waves \cite{BazBar13b}. 

\begin{figure}
\centering
\includegraphics[height=5cm]{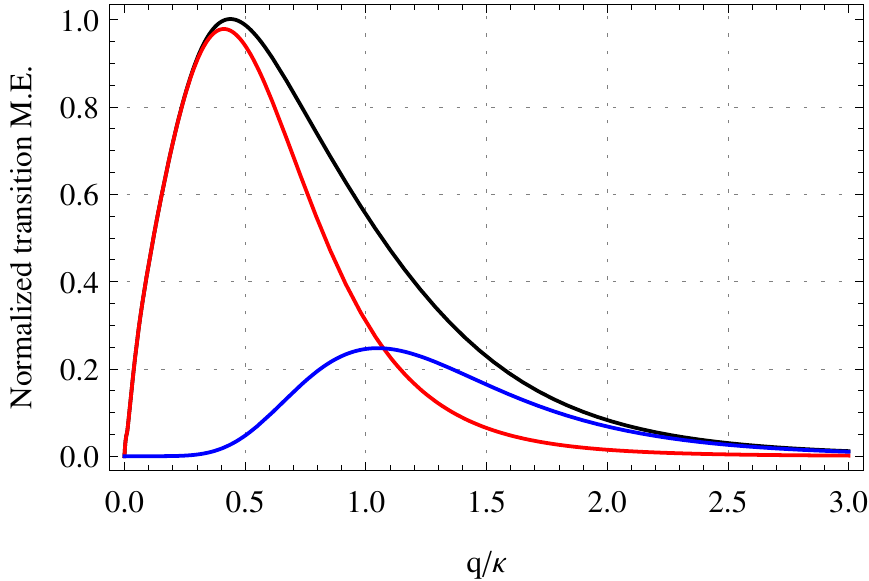}
\caption{\label{fig:ME}  
The normalized three-body transition matrix element, Eq. (\ref{t-frozen}), 
as a function of the relative momentum $q/\kappa$. 
The sum (black), $r^2$ (red, peaked at $q = \kappa/2$), and 
quadrupole (blue, peaked at $q = \kappa$) terms are given for the unitary point,
$|a|=\infty$.}
\end{figure}

The relative contribution of the $s,d$ modes to the trimer formation is 
displayed in  Fig. \ref{fig:ME}, where the last term in parenthesis on 
the rhs of Eq. (\ref{t-frozen}) is presented normalized, along with the 
$s$ and $d$ components. 
Similarly to the dimer formation case \cite{BazLivBar12}, 
the $s$-wave association is peaked around $q=\kappa /2$, 
while the $d$-wave association is peaked at $q=\kappa$. 

%===================
\section{Conclusion}
%===================
We have applied the multipole expansion to study
trimer photoassociation in ultracold atomic gases.
The two dominant modes, at order $k^5$, are studied and their relative 
contribution is shown.
Log periodic oscillations are shown in two cases: (i) for the leading s-wave
mode, near the threshold, and (ii) for all partial waves at the high frequency 
tail.

\begin{acknowledgements}
This work was supported by the ISRAEL SCIENCE FOUNDATION (Grant No.~954/09).
We would like to thank L. Khaykovich and N. Nevo-Dinur 
for their useful comments and suggestions during the preparation 
of this work.
\end{acknowledgements}

% BibTeX users please use
%\bibliographystyle{spbasic}
%\bibliography{../mybib}   

\end{document}